\documentclass[runningheads]{llncs}

\usepackage{color}
\usepackage{soul}
\usepackage{amssymb}
\usepackage[english]{babel}
\usepackage{amsfonts}
\usepackage{amsmath}
\usepackage{physics}
\usepackage{mathtools}
\usepackage{mathrsfs}
\usepackage{cite}
\usepackage{comment}
\usepackage{graphicx}
\usepackage[margin=1.00in]{geometry}

\begin{document}

\title{Privacy-accuracy trade-offs in noisy digital exposure notifications}
\author{Abbas Hammoud\inst{1} \and Yun William Yu\inst{1}}
\authorrunning{A. Hammoud and Y.W. Yu}
\institute{$^1$University of Toronto\\
\email{abbas.hammoud@mail.utoronto.ca, ywyu@math.toronto.edu} }
\maketitle              

\begin{abstract}
Since the global spread of Covid-19 began to overwhelm the attempts of governments to conduct manual contact-tracing, there has been much interest in using the power of mobile phones to automate the contact-tracing process through the development of exposure notification applications. The rough idea is simple: use Bluetooth or other data-exchange technologies to record contacts between users, enable users to report positive diagnoses, and alert users who have been exposed to sick users. Of course, there are many privacy concerns associated with this idea. Much of the work in this area has been concerned with designing mechanisms for tracing contacts and alerting users that do not leak additional information about users beyond the existence of exposure events. However, although designing practical protocols is of crucial importance, it is essential to realize that notifying users about exposure events may itself leak confidential information (e.g. that a particular contact has been diagnosed).
Luckily, while digital contact tracing is a relatively new task, the generic problem of privacy and data disclosure has been studied for decades.
Indeed, the framework of differential privacy further permits provable query privacy by adding random noise.
In this article, we translate two results from statistical privacy and social recommendation algorithms to exposure notification.
We thus prove some naive bounds on the degree to which accuracy must be sacrificed if exposure notification frameworks are to be made more private through the injection of noise.


\end{abstract}

\section{Introduction}
The containment of the Covid-19 pandemic has required the mobilization of the largest global public health response in decades \cite{fisher2020global}.
As such, a wide range of new digital health strategies have been quickly deployed at scale \cite{fagherazzi2020digital}; while this rapid deployment has been necessary given the urgency of the situation, it is still important to responsibly analyze the implications of these strategies \cite{ienca2020responsible}.
We focus in this paper on the privacy implications of digital contact tracing smartphone apps \cite{ferretti2020quantifying}, also known as exposure notification apps \cite{Apple_Google}.

It has been established that the Covid-19 virus is subject to presymptomatic and asymptomatic transmission \cite{Asymptomatic_transmission_1}, posing a serious obstacle for containment because infected individuals can spread the virus without displaying any symptoms.
However, by identifying recent contacts of diagnosed individuals and persuading them to isolate, contact tracing can lower the reproductive number.
Contact-tracing efforts were initially manual; healthcare professionals collect information about recent contacts from patients and alert those contacts about their high-risk status and the next steps to be taken \cite{Manual_contact_tracing}.
Of course, there are some limitations to this, including the inability of patients to report contacts with strangers and the discomfort they may feel when revealing information about contact identities and times to healthcare professionals \cite{Disclosure_discomfort}.
On a larger scale, if infection rates continue to grow, the volume of patients may render manual contact tracing infeasible \cite{ferretti2020quantifying}.
It is here that automated digital contact tracing/exposure notification apps enter the picture \cite{kleinman2020digital, Apple_Google, Pan_European, bay2020bluetrace}, as they promise the ability to notify end-users of potential exposure to Covid-19 while reducing the amount of manual labor required.

By definition, contact tracing apps are designed to track colocation (though generally not directly the location) of app users, which raises significant privacy concerns \cite{bengio2020need}.
As such, a big focus of the design of contact tracing protocols is on privacy \cite{levine1988contact, cho2020contact, raskar2020apps, rivest2020pact, trieu2020epione, troncoso2020decentralized}.
Many of the attempts to characterize and guarantee the privacy of contact-tracing apps have focused on the specific mechanisms they use \cite{Mechanism_Privacy_1, MSB}; an ideal algorithm from a privacy-maximalist perspective leaks no information to any party other than that necessary to inform the end-user of an exposure.
Indeed, the Google-Apple Exposure Notification framework \cite{Apple_Google}, which is currently the predominant protocol adopted by many national tracing apps, is designed to reveal only the minimum information necessary to tell the end-user that they were exposed, as well as some metadata about the individual exposure(s).
This provides nearly no data to 3rd parties (including unfortunately public health officials), but instead is able to provide a strong guarantee of user privacy protection against all but their contacts.

The security analyses for these mechanisms is generally predicated upon hiding everything except for the result of the query (in this case, determining the exposure(s)), usually by making use of cryptography.
However, the result of the query itself leaks information; indeed, in any digital contact tracing app without strong validation of user identities---which is of course difficult when user privacy needs to be protected---it is possible using only logarithmically many identities for a user or other institution (such as a business) to infer details about their own exposure events. By pairing that information with a record of other information about their contacts, this enables revealing the infection status of any of their contacts (or perhaps customers) using the exposure notification system \cite{Inherent_limitations}.
This privacy leak is agnostic to the algorithm used, and is instead `inherent' to digital contact tracing, because it simply requires making repeated valid queries to the contact tracing system.

Fortunately, there exists much work in the scientific literature that has wrestled with controlling the leakage of information from the output of database queries, and we can frame contact tracing as a form of distributed database query on the part of the end-user.
By making use of results from the differential privacy and statistical database privacy literature \cite{differential_privacy}, we may hope to protect against information leakage due to the output of the query.
Unfortunately, there is often some trade-off between utility and privacy, and there is some minimum threshold of added noise needed to guarantee any particular level of privacy, which may reduce the accuracy of contact tracing.

In this paper, we recast the problem of contact tracing as a special case of two other database query settings.
First, a contact-tracing app needs to find out which of a user's contacts have been exposed in order to alert them; this can be thought of as a query for the ones in a binary database containing COVID statuses of the user's contacts. Alternatively, users of a contact-tracing app can be thought of as nodes in a simple graph, with edges denoting contact events with diagnosed users. In this setting, alerting the user can be thought of as recommending one node to another, as in social recommendation algorithms \cite{social_recommendation_survey}.

We thus translate two results from the mathematical privacy literature to inherent limitations of contact-tracing apps.  The first, from statistical databases \cite{dinur_nissim}, is a lower bound on the error of queries for the number of diagnosed contacts in subsets of a user's contact database below which an adversary can almost completely reconstruct the database. The second, from graph-based social recommendation algorithms \cite{GRS}, is an explicit trade-off between the (differential) privacy and accuracy of alerting some target user about their high-risk contacts. We carefully adapt the settings of these results in their original forms to the setting of a contact-tracing app, which will allow us to deduce the conditions under which they apply to contact-tracing. We then use these conditions to re-state the results as inherent privacy/accuracy trade-offs for digital contact tracing. Finally, we use specific numerical examples as the base for a discussion of the results.

\section{Methods}
In this section, we present the results we apply as they appear in their original forms \cite{dinur_nissim, GRS}. Although they can be stated in simpler ways, our goal of translating them over to contact-tracing apps requires that we understand properly the settings in which they are applicable. Therefore, it is more appropriate to include all of the conceptual groundwork on which they are built. 
\subsection{Statistical privacy}
The problem of releasing data without compromising the privacy of its owners has been studied for decades. It is becoming increasingly relevant today as more and more data is collected and published for research and other purposes. Although anonymizing data is a good first step, it has been shown that this is not enough to protect privacy when the data can be linked with external information \cite{sweeney_unique}.

One solution to this is k-anonymity \cite{sweeney_k}: for every subset of the attributes in a database (like address or date of birth), every record in the database is required to be indistinguishable from at least $k-1$ other records. It is known that k-anonymity is still prone to attacks, including linkage, homogeneity, and background knowledge attacks \cite{k_anon_attacks}. Another popular way of protecting privacy is to inject noise into the database. This could include adding noise to the raw data or to answers to queries, depending on the situation. This method has motivated many privacy definitions, the most famous of which is differential privacy \cite{differential_privacy}. The idea behind differential privacy is that the information that can be deduced from the database should not depend on any single record, meaning that it should be the `same' when that record is removed. The results of Dinur and Nissim \cite{dinur_nissim}, which we use here, are actually part of the motivation for the creation of the differential privacy definition.
In \cite{dinur_nissim}, Dinur and Nissim show that subset queries on binary databases are very nonprivate if the errors on query responses are smaller than specific lower bounds. To understand their results, we begin with some definitions. 

\begin{definition}
\label{def1}
A binary database is a set  $D=\{d_{1}, d_{2},...,d_{n}\}$ where $d_{i}\in \{0,1\}$. 
\end{definition}

\begin{definition}
\label{def2}
 A subset query is a subset $q\subset [n]=\{1,...,n\}$, and the \textit{true answer} to the query is the number of $1$'s it contains i.e. $\sum_{i\in q}d_i$.
 \end{definition}
 
\begin{definition}
\label{def3}
 A database algorithm is a function $A:\mathbb{P}(D)\rightarrow \mathbb{R}$ that responds to queries. $A$ is said to be \textit{within $\epsilon$ perturbation} if\\
\[
    |A(q)-\sum_{i\in q}d_{i}|<\epsilon \text{ for all } q\subset [n]
\]
\end{definition}

\begin{definition}
\label{def4}
 We use $dist(D,E)$ to denote the hamming distance between 2 databases $D$ and $E$ of equal length i.e. $dist(D,E)=\textrm{card}(\{i: d_{i}\neq e_{i}\})$.
\end{definition}

Before moving on to the author's results, let us discuss their approach to privacy. They argue that there is no universal definition of privacy that works in every situation. Privacy definitions may depend on the type of query, the assumed background knowledge and ability of the adversary, the nature of the information we want to protect, as well as many other things. However, although it may be difficult to decide what it means for a database algorithm to be private, it is much easier to characterize things that no reasonable privacy definition can allow the database algorithm to do. One of those things is the ability to reconstruct almost all of the database. This is the motivation behind the authors' definition of non-privacy. 

\begin{definition}
\label{def5}
 A database-database algorithm pair $(D,A)$ is said to be $t(n)$-non-private if, for every $\epsilon>0$, there is an algorithm $M$ with access to $A$ and time complexity $t(n)$ that outputs binary databases of the same size $n$ as $D$ and satisfies:
\[
    Pr[M \text{ outputs c such that } dist(c,D)<\epsilon n]~>~1-neg(n)
\]  
where the probability is taken over coin tosses of $A$ and $M$, and $neg(n)$ is a function that decreases faster than any inverse polynomial i.e. for each $c>0$, $neg(n)<n^{-c}$ for sufficiently large $n$.
\end{definition}

Intuitively, for any positive number, we can find an algorithm that outputs, with high probability, a database whose hamming distance from $D$ is smaller than that number. This means that there is a high probability of reconstructing the database with an arbitrarily small error, captured by the hamming distance. The definition certainly captures a situation that no privacy definition should allow. We are now ready to state the result.

\begin{theorem}
\label{thm1}
 Let $D$ be a binary database and $A$ be a database algorithm that is within $o(n)$ perturbation. Then $(D,A)$ is $exp(n)$-nonprivate.
\end{theorem}

\subsection{Graph recommendation algorithms}
The problem of recommending people or items to users based on their online networks is a classic one, and we see examples of it all around us today. These include social media platforms recommending people to each other based on their social networks, online retailers recommending items to users based on what they have already bought, and online streaming services recommending films/series/documentaries to users based on what they have already watched. This problem of social recommendation has been tackled in many ways, including filtering and clustering techniques \cite{SR_1, SR_2, SR_3}. One of the main concerns in social recommendation is protecting the privacy of the network's members \cite{SR_privacy_1}. This problem has been studied, and solutions have been created to address specific privacy limitations \cite{SR_privacy_sol1, SR_privacy_sol2, SR_privacy_sol3}.

We will adapt the work of Machanavajjhala et al.
\cite{GRS}, which examines the privacy of probabilistic social recommendation algorithms that recommend nodes in a graph to some target node based on the target node's connections. These nodes could be people in a social network like Facebook, and the recommendation algorithm could recommend people as potential friends based on their friends' friends. This is the famous ``people you may know'' feature. Another example is a bipartite graph of items and people, as in the Amazon purchase graph, and the goal of the recommendation system is to recommend items to people based on their past purchases and the purchases of other people who have purchased similar items. Of course, such systems come with an inherent privacy risk. Since they use connections (edges) in a graph to make recommendations, they pose the risk of revealing sensitive graph connections in the process of making these recommendations. For example, a system that recommends friends based on your friend's friends could reveal info about that friend's friends list (that the friend may have chosen to hide from everyone). Let's get into the specifics of \cite{GRS}.

We begin with a graph $G=(V,E)$ of $n$ nodes. Letting $r$ denote the target node, we define $u_{i}^{G,r}$ to be the utility of recommending node $i$ to node $r$. We shall stick to the notation $u_{i}$ when the graph and target node are clear from context.
This $(n-1)$-dimensional utility vector is required to be a function only of the graph structure, and is assumed to be given as an input to the problem from external sources \cite{GRS}.
In other words, where the utility vector comes from is irrelevant to the proofs; the authors are only concerned with the conditions it must satisfy. 

A recommendation algorithm is a function that maps an $(n-1)$-dimensional vector of utilities $u^{G,r}$ to an $(n-1)$-dimensional vector of probabilities $p^{G,r}$. Here, $p_{i}^{G,r}$ is the probability of recommending the node $i$ to node $r$. This captures the general case of probabilistic algorithms, and deterministic algorithms can be treated as a special case. For example, the algorithm that recommends the node of highest utility (denoted $R_{best}$) outputs a vector of zeros with a 1 at that node. Of course, such an algorithm injects no noise, and as such does not improve privacy over the baseline. Next, we define the expected utility of each recommendation made by a recommendation algorithm.

\begin{definition}
\label{def6}
 Given a recommendation algorithm $R$ as defined above and a utility vector $u^{G,r}$, the expected utility of the recommendation of $R$ is defined as:

\[
    E_{R}(u^{G,r}) = \sum_{i}u_{i}^{G,r}\cdot p_{i}^{G,r}
\] 
\end{definition}
One can see how this encodes the ``average'' utility of a recommendation: recommending higher utility nodes with a higher probability results in a higher expected utility. Next, we move on to the authors' privacy definition; they adapt the differential privacy of Dwork [2] to the graph setting by defining databases that differ by one element as graphs that differ by one edge. 

\begin{definition}
\label{def7}
 A recommendation algorithm $R$ satisfies $\epsilon$-differential privacy if for any pair of graphs $G$ and $G'$ that differ in only one edge and every set of possible recommendations $S$, \\
\[
    Pr[R(G)\in S]~\leq ~ e^{\epsilon}~Pr[R(G')\in S]
\]
\end{definition}
This privacy definition is a guarantee to protect the connections between nodes. This will be important in the next section. Next, let's define accuracy.

\begin{definition}
\label{def8}
 The accuracy of an algorithm $R$ is defined as:\\
\[
   A(R)= \min_{u}\frac{\sum_{i}u_{i}p_{i}}{u_{max}}
\]  
\end{definition}
In \cite{GRS}, the minimum above is described to be taken over all possible input vectors to the algorithm. In other words, if the accuracy is $1-\delta$, then $\frac{\sum_{i}u_{i}p_{i}}{u_{max}}\geq 1-\delta$ for every utility vector and corresponding probability vector. Moreover, there is a utility vector for which the equality is achieved. Note that this accuracy is a lower bound on the ratio of expected utility to maximum utility; therefore, it can be thought of as an approximation of utility. In fact, the main result of the paper is a privacy/accuracy trade-off. Let us now discuss the 3 axioms that the authors require to be true for their results to hold.

\textbf{Axiom 1 (Exchangeability)}: If $G$ is a graph and $h$ is an isomorphism on the nodes that fixes the target node $r$ and we denote $h(G)$ by $G_{h}$, then $u_{i}^{G,r}=u_{h(i)}^{G_{h},r}$ for all nodes $i\neq r$.

Notice that the axiom deals with a classic graph isomorphism i.e. an isomorphism of nodes that does not mess with connections. The idea of Axiom 1 is that a non-target node $i$ shares the same connection with $r$ in $G$ as does $h(i)$ with $h(r)=r$ in $G_h$, and that this forces the utility of recommending $i$ to $r$ to be the same as that of recommending $h(i)$ to $r$. This formalizes the idea that the utilities should only depend on the structural properties of the graph, as opposed to the identities of the nodes. This will have massive implications when we try to translate the results of \cite{GRS} to statements about contact-tracing apps in the next section. 

\textbf{Axiom 2 (Concentration)}: There exists $S\subset V(G)$, with $|S|=\beta$, satisfying $\sum_{i\in S}u_{i}\geq \Omega(1)\sum_{i\in V(G)}u_{i}$.

This axiom is saying that there is a group of nodes whose combined utility is at least a constant multiple of the total utility in the graph. This is effectively saying that most of the utility in the graph should be concentrated in a small subset of the nodes. For example, by any reasonable and practical definition of utility, how many people have a high utility of being recommended as friends to the average Facebook user?

\textbf{Axiom 3 (Monotonicity)}: Our algorithm is monotonic in the traditional sense; namely, $u_{i}>u_{j}$ implies that $p_{i}>p_{j}$ for every distinct pair of non-target nodes $i,j$.

Although the recommendation algorithm should not recommend high utility nodes with very high probabilities in the interest of privacy, it is reasonable to require that it recommends higher utility nodes with a higher probability. There is some more notation we need before stating the main results of \cite{GRS}.  

Let $L$ be the set of lowest utility non-target nodes and $H$ be the set of highest utility non-target nodes. Notice that any graph can be modified by adding and/or deleting edges to turn a node in $L$ into a node in $H$. This is easily shown to be true through a simple visualization. Let $n_{l}\in L$ be the node whose utility we would like to elevate. We can imagine adding/deleting a number of edges to make the connections of $n_{l}$ identical to those of one of the nodes in $H$, which we call  $n_{h}$. Once this is done, the graph isomorphism that maps $n_{l}$ to $n_{h}$ and fixes everything else guarantees that the modified $n_{l}$ has the same utility as $n_{h}$ by Axiom 1. We denote by $t$ the number of edges that must be added and/or deleted to achieve this.

Finally, let $c\in (0,1)$; we can use this parameter to split nodes into a high and low utility group. Let $V_{hi}^{r}$ be the set of nodes $1,...,k$ which each have a utility strictly greater than $(1-c)u_{max}$. Let $V_{lo}^{r}$ be the set that contains the nodes $k+1,...,n$ which each have utility at most $(1-c)u_{max}$. We can now state the general bounds of \cite{GRS}.

\begin{lemma}
\label{lem1}
If $R$ is a recommendation algorithm that is $\epsilon$-differentially private, has accuracy $(1-\delta)$, and satisfies Axioms 1-3, then:\\
\[
    \frac{1}{t}\left[\ln\left(\frac{c-\delta}{\delta}\right)+\ln\left(\frac{n-k}{k}\right)\right]~\leq ~\epsilon
\]  
\end{lemma}

Notice that the expression above can be rearranged from a lower bound of the privacy parameter $\epsilon$ to an upper bound of the accuracy parameter $1-\delta$. The authors use Lemma 1 to prove the main result of the paper:

\begin{theorem}
\label{thm2}
 For a graph with maximum degree $d_{max}=\alpha \log n$, a differentially private agorithm can guarantee constant accuracy (approximation to utility) only if
\[
    \epsilon ~\geq \frac{\log{n}-\log{\beta}-\log{\log n}}{4d_{max}}=~\frac{1}{\alpha}\left(\frac{1}{4}-o(1)\right)
\]\\
where $\beta$ is from Axiom 2.
\end{theorem}

\section{Results}
In this section, we connect the results of the previous section to contact-tracing apps. We begin by looking at the conditions under which the results apply to contact tracing. This is straightforward for the subset-sum queries of section 2.1, but takes a bit more work for graph recommender systems. The main reason for this is the set of axioms that the utility functions of section 2.2 must obey. Although these conditions are easy to understand in the abstract, we must understand what they mean for contact-tracing apps. Once this is done, we will re-state the results as they apply to the apps. In the next section, we give concrete examples of their applications, which will set the stage for our discussion about their implications to digital contact-tracing. 

\subsection{Statistical privacy}
To begin, we will understand the binary databases of Definition \ref{def1} to contain the Covid-19 statuses of a group of app users. Furthermore, an adversary of a database is anyone who can conduct subset queries on it. In this case, any user can serve as an adversary seeking to find out the Covid-19 status of their contacts, if they are able to conduct subset queries. In practice, all this requires is that the app inform the user of the number of positive contacts (the number of contacts that involved one or both users reporting a positive diagnosis to the app in some specified time interval around the contact event) they have had in some period of time. The user can then conduct subset queries on a database by coming in `contact' with specific subsets of the database---in a similar to procedure to that given in \cite{Inherent_limitations}, synthetic identities or additional queries will both suffice to allow multiple such subset-sum queries. If the user chooses a group whose members they can come in contact with, they can use the result of Section 2.1 to reconstruct the binary database containing the group's statuses. Now, Theorem \ref{thm1} in Section 2.1 only tells us that this reconstruction is possible, but it does not provide a recipe for actually reconstructing anything. This recipe is provided by Dinur and Nissim's proof of the Theorem in \cite{dinur_nissim}. We have chosen not to include it in the interest of clarity, but we will mention aspects of it that are important to understanding the implications of the Theorem to digital contact-tracing. 

In the proof of Theorem 1, Dinur and Nissim provide a brute force algorithm that reconstructs a binary database $D$ of length $n$ using a database algorithm that is within $o(n)$ perturbation. Recall that this means there is some $o(n)$ function, call it $f$, that is greater than the error of the database algorithm for any subset query. The reconstruction algorithm has a $100\%$ chance of finding a binary database of length $n$ whose hamming distance from $D$ is at most $4f$.Because it naively performs a brute force search, it may require up to $2^{2n}$ subset queries, so it has exponential time complexity. These features of the reconstruction algorithm allow it to satisfy exponential non-privacy; however, they are more specific than the statement of the non-privacy definition, so we'll use them to translate Theorem 1 to a result about contact tracing apps.

\begin{theorem}
\label{thm3} 
Suppose a contact-tracing app allows a user to conduct subset queries on a binary database $D$ of length $n$, and the subset queries are within some $f(n)=o(n)$ perturbation. The user can, with at most $2^{2n}$ subset queries, find a binary database of length $n$ that differs from $D$ in at most $4f$ entries.

\end{theorem}

\subsection{Graph recommendation algorithms}
As mentioned earlier, there is a bit more work involved in translating Theorem \ref{thm2} into a statement about contact-tracing apps. This is because of the three axioms that \cite{GRS} assumes about the utility function of the graph, which we need to adapt into reasonable contact tracing axioms.
First, we contextualize this adaptation by re-imagining the users of a contact tracing app as a graph. We will assume that our discussion here concerns some specific interval of time $I$. This means that our user base will consist of a group of users who had the app downloaded at any point during $I$. Moreover, we will concern ourselves with contact events that occurred in $I$. With that, we can think about defining our graph. It may seem intuitive to use nodes to represent users and edges to represent contacts. However, we will see below when we attempt to adapt Axiom 1 that this edge definition is flawed. 
Instead, we begin by redefining a ``recommendation'' to better suit the setting of a contact tracing app. Next, we discuss utility and adapt Axiom 1, which will inform our edge definition. 

In this setting, ``recommendations'' are just alert messages about the risk posed to a target node by other nodes. Intuitive as that may sound, there is a clear inconsistency in the use of the word ``recommendation'' here: a privacy-conscious contact-tracing app will never recommend a non-target node to the target node in as explicit a way as Facebook's ``People you may know" or Amazon's ``Customer's who bought this item also bought" features. So, how do we reconcile this with our attempt to use the results of \cite{GRS}? We can easily redefine the recommendation of a non-target node to a target node as an alert message, sent to the target node, that does not explicitly identify and is in some way motivated by the non-target node. The simplest motivation for such an alert message is a direct contact with a sick non-target node. The reason this redefinition is possible is that \cite{GRS} does not make any assumptions about the specific mechanism of a recommendation. 

We now move on to the crucial topic of utility. Although the authors of \cite{GRS} assume that the utility vector specifying the utility of recommending each non-target node to the target node is given, we cannot get away with just that. The reason for this is that the axioms of \cite{GRS} specify assumptions about utilities. These assumptions place natural restrictions on the kinds of utilities that the results of \cite{GRS} can be used to talk about, so it is important to characterize these restrictions in order to understand the conditions under which Theorem \ref{thm2} applies to contact-tracing apps. The adaptation of Axiom 1 will allow us to do this.

We begin this adaptation by discussing why the definition of an edge as a contact event between nodes is flawed. This flaw arises from the fact that such a definition does not allow us to satisfy Axiom 1. We clarify this by looking at a basic example in which this definition is adopted. Suppose our graph is made up of 3 nodes $\{n_{1},n_{2},n_{3}\}$. Let's designate $n_{1}$ as the target node and assume that the only edges in the graph are $(n_{1},n_{2})$ and $(n_{1},n_{3})$. Now, suppose that $n_{2}$ becomes sick. Although we do not have an explicit definition of utility, let's take it for granted that there is much greater utility in recommending $n_{2}$ than $n_{3}$ i.e. $u_{2}^{G,1}>u_{3}^{G,1}$. Now, consider the second most trivial $n_{1}$-fixing isomorphism of the graph: the one that maps $n_{1}$ to itself, $n_{2}$ to $n_{3}$, and $n_{3}$ to $n_{2}$. Let us call it $h$. It is clear that $u_{2}^{G,1}\neq u_{3}^{G,1}=u_{h(2)}^{G_{h},1}$. This is is a violation of Axiom 1.

This violation makes sense since the edges of our graph only represent contacts between users, which fails to take into account that only sick contacts will be recommended to the user. In other words, there is information external to the structural properties of the graph that impacts the utility. A simple way of fixing this is to redefine edges in our graph. We can define two nodes as sharing an edge if the nodes have been in contact and one of the nodes reported a positive diagnosis to the app in some interval of time around the contact event. Although this fixes the example above by removing the edge between $n_{1}$ and $n_{3}$, the problem has far reaching consequences because Axiom 1 forces us to only examine utilities that can be determined completely by the structural properties of a graph. Since the relationships captured by a graph are fundamentally binary ones, this places a natural limit on the complexity of the information that can be used to determine the utility functions we analyze using \cite{GRS}. 

The above makes it clear that we must define edges according to the variables that will determine the utility vector. To clarify this, we consider the simplest way of doing it, which is to specify a set of binary questions $\{k_{1},k_{2},...,k_{m}\}$ about a pair of nodes and to place an edge between them if the answers to all of the questions are "Yes". These questions represent the binary variables that utility will take into account. We have already given examples of $k_{i}$: 
"have the two users been in contact in the time period that the graph represents" and "has one of the two users reported a positive diagnosis to the app in a time interval around the contact event". One could generate $m$-bit strings for each pair of users and place edges between users who have a string of $1$'s. This is effectively how most current contact-tracing efforts get the job done. A sick person is asked to recall their contacts, and those contacts are alerted. In that case, $m=2$, with $k_{1}$ and $k_{2}$ being the two questions given above.

To summarize the discussion up to this point, we've used Axiom 1 to better understand the restricted set of utilities to which the results of \cite{GRS} can be applied i.e. those that are completely determined by the structural properties of the graph. Moreover, this informed our definition of edges by forcing it to take into account the variables that determine utility. Let us move on to the adaptations of Axioms 2 and 3.

The statement of Axiom 2 is that most of the utility in the graph is contained in a few nodes. This is a reasonable assumption in our setting, since the subset of all users that the average target node will come in contact with in the time interval $I$ is tiny compared to the whole user base. Moreover, only a fraction of contacts will become sick. Of course, this assumes that reasonably many people have downloaded the app. Luckily, that is one of the prerequisites for the success of any contact-tracing app, which makes the assumption a natural one to make.

When considering utility related to the alerting process, the assumption of axiom 3 is not only a very reasonable one to make, but also seems like a requirement of the app. However, there may be utility functions we have not investigated that warrant violating this axiom in the interest of privacy. The implication of this is a restriction similar to the one from our discussion of Axiom 1: we can only use \cite{GRS} to talk about recommendation algorithms that are monotonic according to Axiom 3. 

Now that we understand what it means for the graph and utility function of a contact-tracing app to satisfy Axioms 1-3, we can restate Theorem \ref{thm3}.

\begin{theorem}
\label{thm4}
Suppose that the graph, utility function, and recommendation algorithm of a contact-tracing algorithm satisfy Axioms 1-3 as discussed above. If the recommendation algorithm is differentially private in the sense of definition 7 and it guarantees constant accuracy, then\\
\[
    \epsilon ~\geq \frac{\log{n}-\log{\beta}-\log{\log n}}{4d_{max}}
\]\\
where $\beta$ is from Axiom 2 and $d_{max}$ is the maximum degree of the graph.
\end{theorem}
\section{Discussion}

In this section, we discuss and look at specific examples of the theorems in the previous one. This will help us better understand their implications for digital contact-tracing.
\subsection{Statistical privacy}
Theorem \ref{thm3} is a trade-off between the the error of subset queries and the accuracy with which an adversary can reconstruct the database. If we set $n=50$ (because a user has fewer than 50 contacts in a day, for example) and the perturbation of subset queries to $3$, then an adversary can reconstruct the database with at most $12$ incorrect entries. This means that the COVID statuses of at least $38$ people are leaked. Although this is a clear privacy limitation of any contact-tracing app that allows the user to conduct subset queries, recall that Theorem \ref{thm3} gives an exponential ($2^{2n}$) upper bound on the of subset queries it would take to perform the reconstruction. This means that Theorem \ref{thm3} is best suited for small databases of less than $10$ users.

Although this is a weakness of the Theorem, there are few things that make up for it. Note first that the naive approach of splitting a large database into many smaller databases---imagine a user carrying around multiple phones for different time periods, or just using a modified version of an app that simulates the same---does not work. Although the reconstruction algorithm can then be run on each of the smaller databases, splitting up the databases does not change the error of each subset query. In our example of the 50 user database above, we can split the database into 10 smaller databases with 5 entries each, so the reconstruction of the original would take at most $10(2^{10})=10240$ subset queries, a far cry from the $2^{100}$ upper bound if the database were not split. However, the problem with this is that the subset queries would still have a perturbation of $3$, meaning the upper bound on the error of each reconstruction is still at most $12$, greater than the size of the database, and so a meaningless bound.

However, this exponential-time limitation on Theorem \ref{thm3} is due to the original reconstruction algorithm brute forcing all possible subsets; in the same paper, Dinur and Nissim also provide a less accurate polynomial-time algorithm \cite{dinur_nissim}. In this paper, we have analyzed the exponential-time algorithm as the error bounds are much easier to compute, but in practice, it is likely that more efficient algorithms exist---indeed, in the limiting case of no error, full reconstruction is possible in logarithmic queries \cite{Inherent_limitations}. Thus, while Theorem \ref{thm3} is not especially strong as a bound, it does still point to the feasibility of translating results from the literature to contact tracing privacy.

\subsection{Graph recommendation algorithms}
Let's talk about Theorem \ref{thm4}. There is a bit more ambiguity in interpreting this Theorem because it is a statement about the $\epsilon$ parameter of edge differential privacy in a graph. This makes the result more difficult to visualize than that of Theorem \ref{thm3}. Consider a specific numerical example: let's assume we are looking at a graph corresponding to a time period of 2 weeks. Although it is easy to choose the database size $n$, it is less clear how we should choose $\beta$ and $d_{max}$.
$\beta$ (from Axiom 2) is 
the number of non-target nodes whose utility of being recommended to the target is greater the total utility in the graph, which means that $\beta$ depends on the utility function and the target node. Based on the intuition we developed for utility and edges in the previous section, a reasonable guess for $\beta$ is the number of positive contacts the target node has had in the 2 weeks. 

Reasoning similarly, a good guess for $d_{max}$ would be the maximum number of positive contact events that any user has had in the 2 weeks. So, if we take a community of $n=1000$ people and consider an average target node, taking $\beta=5$ and $d_{max}=20$ seems reasonable. According to Theorem \ref{thm4}, this means that $\epsilon>0.02$. Since we have not talked about ways of achieving differential privacy in a graph of users of a contact-tracing app, we are not in a position to interpret the exact significance of such a lower bound on the $\epsilon$ parameter. Indeed, from a practical point of view, applying differential privacy to decentralized contact tracing apps runs into another problem, which is that these bounds measure information leakage from single queries; differential privacy bounds are composable when multiple queries are performed, but that means that when we cannot strongly limit the number of queries the adversary performs, these bounds are much less applicable.

However, we can see that, should the designers of a contact-tracing app figure out a way to inject noise that endows the graph with edge differential privacy but maintains a constant accuracy of the recommendation algorithm while rate-limiting the number of queries users can perform, there is a lower bound on the $\epsilon$ parameter. This is important to keep in mind because we would like this parameter to be as small as possible. Moreover, a trade-off between the noise injected and this lower bound is specified by Lemma \ref{lem1}. We can see there that a lower $\delta$ (corresponding to less noise and a higher accuracy/utility $1-\delta$) means a higher lower bound, which in turn means a lower degree of (differential) privacy.

\section{Conclusion}
Although the bounds we present here in this paper may not seem particularly strong at a first glance, they still concretely establish the trade-off in privacy and utility when it comes to adding noise to digital contact tracing results.
Indeed, at the moment, deployed contact tracing apps, while designed to maximize privacy when it comes to everything except the query results, err on the side of as much accuracy as possible for the query result itself; i.e. no noise is ever deliberately injected.
This is of course a philosophical question: Is it ever OK for an exposure notification app to lie to users about a potential exposure in the interest of privacy?
In the subset-sum queries, a perturbation of even just `1' means that a large fraction of users who are exposed will not be informed.
Even though noise injection can be done in a nuanced way to omit false positives, and there may be more sophisticated mechanisms for differential privacy, this question is beyond the scope of this paper, and is indeed a question for healthcare professionals and policy makers, rather than algorithms designers.

We believe that these kinds of limitations are relevant to everybody involved in the process of designing and deploying a contact-tracing app.
For example, healthcare professionals and policy makers might want to communicate them to the public in order to increase transparency, which is important in trying to guarantee widespread adoption of a digital contact tracing app.
Moreover, since these limitations exist without any assumption on the specific mechanisms used by the app, app designers should know about them to avoid wasting time and resources in attempts to ameliorate them through app design.

In this paper, we have seen how privacy results from statistical databases and social recommendation algorithms can be adapted to statements about contact-tracing apps that do not depend on the specific mechanisms that the apps use.
In our work, these statements have been mathematically precise privacy-utility trade-offs.
Even though these initial bounds are not a strong or tight as could be, they already reveal some of the privacy limitations that policy makers will have to work with when deploying digital contact tracing frameworks: it is not possible to provide perfect privacy while also maximizing accuracy and utility.
We hope (1) that future investigations will produce stronger bounds on the privacy limitations of contact tracing queries, and (2) that these investigations will help inform policy makers when making decisions about how to balance public health needs with individual privacy.

\newpage
\bibliography{References.bib}{}

\begin{thebibliography}{10}

\bibitem{fisher2020global}
D.~Fisher and A.~Wilder-Smith, ``The global community needs to swiftly ramp up
  the response to contain covid-19,'' {\em The Lancet}, vol.~395, no.~10230,
  pp.~1109--1110, 2020.

\bibitem{fagherazzi2020digital}
G.~Fagherazzi, C.~Goetzinger, M.~A. Rashid, G.~A. Aguayo, and L.~Huiart,
  ``Digital health strategies to fight covid-19 worldwide: challenges,
  recommendations, and a call for papers,'' {\em Journal of Medical Internet
  Research}, vol.~22, no.~6, p.~e19284, 2020.

\bibitem{ienca2020responsible}
M.~Ienca and E.~Vayena, ``On the responsible use of digital data to tackle the
  covid-19 pandemic,'' {\em Nature medicine}, vol.~26, no.~4, pp.~463--464,
  2020.

\bibitem{ferretti2020quantifying}
L.~Ferretti, C.~Wymant, M.~Kendall, L.~Zhao, A.~Nurtay, L.~Abeler-D{\"o}rner,
  M.~Parker, D.~Bonsall, and C.~Fraser, ``Quantifying sars-cov-2 transmission
  suggests epidemic control with digital contact tracing,'' {\em Science},
  vol.~368, no.~6491, 2020.

\bibitem{Apple_Google}
Apple and Googe, ``Privacy-preserving contact tracing,'' 2020.

\bibitem{Asymptomatic_transmission_1}
Z.~Hu, C.~Song, C.~Xu, G.~Jin, Y.~Chen, X.~Xu, H.~Ma, W.~Chen, Y.~Lin,
  Y.~Zheng, J.~Wang, z.~hu, Y.~Yi, and H.~Shen, ``Clinical characteristics of
  24 asymptomatic infections with covid-19 screened among close contacts in
  nanjing, china,'' {\em medRxiv}, 2020.

\bibitem{Manual_contact_tracing}
J.~Flint, S.~Burton, J.~Macey, S.~Deeks, T.~Tam, A.~King, M.~Bodie-Collins,
  M.~Naus, D.~MacDonald, C.~McIntyre, {\em et~al.}, ``Assessment of in-flight
  transmission of sars--results of contact tracing, canada.,'' {\em Canada
  communicable disease report= Releve des maladies transmissibles au Canada},
  vol.~29, no.~12, p.~105, 2003.

\bibitem{Disclosure_discomfort}
W.~F. Flanagan, ``Equality rights for people with {AIDS}: Mandatory reporting
  of {HIV} infection and contact tracing,'' {\em McGill LJ}, vol.~34, p.~530,
  1988.

\bibitem{kleinman2020digital}
R.~A. Kleinman and C.~Merkel, ``Digital contact tracing for covid-19,'' {\em
  CMAJ}, vol.~192, no.~24, pp.~E653--E656, 2020.

\bibitem{Pan_European}
``Pan-european privacy-preserving proximity tracing,'' 2020.

\bibitem{bay2020bluetrace}
J.~Bay, J.~Kek, A.~Tan, C.~S. Hau, L.~Yongquan, J.~Tan, and T.~A. Quy,
  ``Bluetrace: A privacy-preserving protocol for community-driven contact
  tracing across borders,'' {\em Government Technology Agency-Singapore, Tech.
  Rep}, 2020.

\bibitem{bengio2020need}
Y.~Bengio, R.~Janda, Y.~W. Yu, D.~Ippolito, M.~Jarvie, D.~Pilat, B.~Struck,
  S.~Krastev, and A.~Sharma, ``The need for privacy with public digital contact
  tracing during the covid-19 pandemic,'' {\em The Lancet Digital Health},
  2020.

\bibitem{levine1988contact}
M.~L. Levine, ``Contact tracing for hiv infection: a plea for privacy,'' {\em
  Colum. Hum. Rts. L. Rev.}, vol.~20, p.~157, 1988.

\bibitem{cho2020contact}
H.~Cho, D.~Ippolito, and Y.~W. Yu, ``Contact tracing mobile apps for covid-19:
  Privacy considerations and related trade-offs,'' {\em arXiv preprint
  arXiv:2003.11511}, 2020.

\bibitem{raskar2020apps}
R.~Raskar, I.~Schunemann, R.~Barbar, K.~Vilcans, J.~Gray, P.~Vepakomma,
  S.~Kapa, A.~Nuzzo, R.~Gupta, A.~Berke, {\em et~al.}, ``Apps gone rogue:
  Maintaining personal privacy in an epidemic,'' {\em arXiv preprint
  arXiv:2003.08567}, 2020.

\bibitem{rivest2020pact}
R.~L. Rivest, D.~Weitzner, L.~Ivers, I.~Soibelman, and M.~Zissman, ``Pact:
  Private automated contact tracing,'' 2020.

\bibitem{trieu2020epione}
N.~Trieu, K.~Shehata, P.~Saxena, R.~Shokri, and D.~Song, ``Epione: Lightweight
  contact tracing with strong privacy,'' {\em arXiv preprint arXiv:2004.13293},
  2020.

\bibitem{troncoso2020decentralized}
C.~Troncoso, M.~Payer, J.-P. Hubaux, M.~Salathe, J.~Larus, E.~Bugnion,
  W.~Lueks, T.~Stadler, A.~Pyreglis, D.~Antonioli, L.~Barman, S.~Chatel,
  K.~Paterson, S.~Capkun, D.~Basin, J.~Beutel, D.~Jackson, B.~Preneel,
  N.~Smart, D.~Singelee, A.~Abidin, S.~Guerses, M.~Veale, C.~Cremers, R.~Binns,
  and C.~Cattuto, ``Decentralized privacy-preserving proximity tracing,'' May
  2020.

\bibitem{Mechanism_Privacy_1}
J.~Chan, S.~Gollakota, E.~Horvitz, J.~Jaeger, S.~Kakade, T.~Kohno, J.~Langford,
  J.~Larson, S.~Singanamalla, J.~Sunshine, {\em et~al.}, ``Pact: Privacy
  sensitive protocols and mechanisms for mobile contact tracing,'' {\em arXiv
  preprint arXiv:2004.03544}, 2020.

\bibitem{MSB}
``Bluetrace: A privacy-preserving protocol for community-driven contact tracing
  across borders,'' 2020.

\bibitem{Inherent_limitations}
Y.~Bengio, D.~Ippolito, R.~Janda, M.~Jarvie, B.~Prud'homme, J.-F. Rousseau,
  A.~Sharma, and Y.~W. Yu, ``Inherent privacy limitations of decentralized
  contact tracing apps,'' {\em Journal of the American Medical Informatics
  Association}, 2020.

\bibitem{differential_privacy}
C.~Dwork, ``Differential privacy: A survey of results,'' in {\em International
  conference on theory and applications of models of computation}, pp.~1--19,
  Springer, 2008.

\bibitem{social_recommendation_survey}
J.~Tang, X.~Hu, and H.~Liu, ``Social recommendation: a review,'' {\em Social
  Network Analysis and Mining}, vol.~3, no.~4, pp.~1113--1133, 2013.

\bibitem{dinur_nissim}
I.~Dinur and K.~Nissim, ``Revealing information while preserving privacy,'' in
  {\em Proceedings of the twenty-second ACM SIGMOD-SIGACT-SIGART symposium on
  Principles of database systems}, pp.~202--210, 2003.

\bibitem{GRS}
A.~Machanavajjhala, A.~Korolova, and A.~D. Sarma, ``Personalized social
  recommendations: Accurate or private,'' {\em Proc. VLDB Endow.}, vol.~4,
  p.~440–450, Apr. 2011.

\bibitem{sweeney_unique}
L.~Sweeney, ``Uniqueness of simple demographics in the us population,'' {\em
  LIDAP-WP4, 2000}, 2000.

\bibitem{sweeney_k}
L.~Sweeney, ``k-anonymity: A model for protecting privacy,'' {\em International
  Journal of Uncertainty, Fuzziness and Knowledge-Based Systems}, vol.~10,
  no.~05, pp.~557--570, 2002.

\bibitem{k_anon_attacks}
A.~Machanavajjhala, D.~Kifer, J.~Gehrke, and M.~Venkitasubramaniam,
  ``l-diversity: Privacy beyond k-anonymity,'' {\em ACM Transactions on
  Knowledge Discovery from Data (TKDD)}, vol.~1, no.~1, pp.~3--es, 2007.

\bibitem{SR_1}
X.~Yang, Y.~Guo, Y.~Liu, and H.~Steck, ``A survey of collaborative filtering
  based social recommender systems,'' {\em Computer communications}, vol.~41,
  pp.~1--10, 2014.

\bibitem{SR_2}
Y.~Yu and X.~Chen, ``A survey of point-of-interest recommendation in
  location-based social networks,'' in {\em Workshops at the Twenty-Ninth AAAI
  Conference on Artificial Intelligence}, 2015.

\bibitem{SR_3}
J.~Bao, Y.~Zheng, D.~Wilkie, and M.~Mokbel, ``Recommendations in location-based
  social networks: a survey,'' {\em GeoInformatica}, vol.~19, no.~3,
  pp.~525--565, 2015.

\bibitem{SR_privacy_1}
V.~Nikolaenko, S.~Ioannidis, U.~Weinsberg, M.~Joye, N.~Taft, and D.~Boneh,
  ``Privacy-preserving matrix factorization,'' in {\em Proceedings of the 2013
  ACM SIGSAC conference on Computer \& communications security}, pp.~801--812,
  2013.

\bibitem{SR_privacy_sol1}
Z.~Jorgensen and T.~Yu, ``A privacy-preserving framework for personalized,
  social recommendations.,'' in {\em EDBT}, vol.~582, 2014.

\bibitem{SR_privacy_sol2}
K.~Liu and E.~Terzi, ``A framework for computing the privacy scores of users in
  online social networks,'' {\em ACM Transactions on Knowledge Discovery from
  Data (TKDD)}, vol.~5, no.~1, pp.~1--30, 2010.

\bibitem{SR_privacy_sol3}
T.~R. Hoens, M.~Blanton, and N.~V. Chawla, ``A private and reliable
  recommendation system for social networks,'' in {\em 2010 IEEE Second
  International Conference on Social Computing}, pp.~816--825, IEEE, 2010.

\end{thebibliography}
\bibliographystyle{ieeetr}

\end{document}